# "La Ola-MJO": a public-friendly nickname for the Madden-Julian Oscillation

Takeshi Izumo[a], Bastien Pagli[a], Claire Rocuet[b], Mayalen Zubia[b], Marania Hopuare[c], Vateanui Sansine[a], Maxime Colin[d]

[a]IRD (French national Institute of Research for sustainable Development), UMR241 SECOPOL laboratory (IRD/UPF/IFREMER/ILM), Tahiti, French Polynesia
[b]Université de Polynésie Française (UPF), UMR241 SECOPOL laboratory (IRD/UPF/IFREMER/ILM), Tahiti, French Polynesia
[c]GEPASUD laboratory, Université de Polynésie Française (UPF), Tahiti, French Polynesia
[d]Leibniz Centre for Tropical Marine Research (ZMT), Bremen, Germany

**The Madden−Julian Oscillation (MJO) is the leading intraseasonal driver of tropical weather, comparable to the interannual "El Niño/La Niña", yet remains largely unfamiliar outside the scientific community. The MJO is atmospherically like a "La Niña" going around the earth, and metaphorically as "La Ola"—Spanish for "The Wave", evoking a stadium wave. Thus a pedagogical nickname such as "La Ola-MJO" could enhance forecast uptake, especially among climate-vulnerable communities.**

The Madden-Julian Oscillation (MJO) rivals the El Niño-Southern Oscillation (ENSO) in its importance for weather and climate intraseasonal variability and extremes (cf. Figs 1 and 2). Its strength is projected to increase in a warming climate, especially in terms of precipitation, and will possibly become more predictable[1,2,3]. It is thus crucial for our society to better take into account the MJO forecasts so that fragile communities can be better prepared, and to adapt to climate change impacts. Yet, although the term MJO is somewhat known by the public in some highly impacted regions such as Australia or Indonesia, it has yet to gain widespread recognition outside the scientific community, conversely to El Niño/La Niña. This is probably because of its long, complicated scientific name. While "El Niño" and "La Niña" have become household terms (in many affected countries such as Australia) that successfully bridge the gap between scientific research and public understanding, the MJO lacks an accessible name that resonates with non-specialists. Piggybacking on "El Niño" and "La Niña", we suggest "La Ola-MJO" as a possible nickname for the MJO, a public-friendly term whose metaphor is pedagogical and easy to remember. Thus, transforming how this critical phenomenon is communicated to the public, media, and stakeholders, while maintaining the connection to Madden and Julian's foundational discovery. "La Ola" means "The Wave" in Spanish (distinct from the term "onda" that also exists in Spanish, e.g. 'ondas ecuatoriales'), and is figuratively used for the 'human wave' that circum-propagates around sports stadium. "La Ola" evokes the two main MJO physical features, an irregularly-oscillating tropical atmospheric wave circum-propagating around the earth (Fig. 1), while phonetically reminding us of "La Niña", its physically-closest well-known climate analog, qualitatively similar in terms of precipitation and wind to the MJO in its main active phase over the Maritime continent (Fig. 2). In essence, we could pedagogically synthesize the MJO as a tropical perturbation, atmospherically like a "La Niña", but that goes around the earth metaphorically as an "Ola", eastward along the equator in about one to two months.

## Why the MJO needs a public-friendly nickname

The MJO is the dominant mode of intraseasonal atmospheric variability in the tropics, with far-reaching impacts on rainfall patterns, monsoons, tropical cyclone activity, and extreme weather/oceanic events such as humid/marine heatwaves[4,5] across the globe[6,7]. The MJO is projected to increase in strength and predictability[2]. Despite advances in MJO prediction skill over the past two decades[8], this knowledge does not sufficiently reach those who could benefit most: farmers planning planting schedules, water managers anticipating flood risks, and communities preparing for extreme events in the climate change fragilizing context.

The term "Madden-Julian Oscillation" presents several barriers to public communication:

- **Technical terminology**: "Oscillation" is abstract and requires scientific background to understand
- **Lack of cultural connection**: Unlike El Niño (named by Peruvian fishermen for its Christmas-season appearance), the MJO has no cultural or regional anchoring
- **Acronym dependency**: While "MJO" is efficient for scientists, acronyms are notoriously difficult for public memory and media adoption, and furthermore can vary among languages (e.g. “OMJ” in Latin languages such as French and Spanish)

This communication gap has real consequences. Subseasonal forecasts (2-4 weeks ahead) that leverage MJO predictions could provide crucial lead time for disaster preparedness and agricultural decision-making, yet this information remains largely inaccessible to end-users who do not understand what the MJO is or why it matters to them.

## The physical connection: “La Niña” and “La Ola”

The nickname “La Ola” – that we propose to add to the MJO acronym – builds on an important conceptual bridge to “La Niña”. During the MJO's main active phase over the Maritime Continent, the patterns of enhanced atmospheric convection and circulation bear a striking qualitative resemblance to “La Niña” atmospheric conditions (Figure 2). Both phenomena feature:

- Intensified convection/precipitation over the Indo-Pacific warm pool (with amplitudes of the same order, with “La Niña” ones being somewhat stronger and more extended)
- Strengthened Walker Circulation with westerly anomalies along the equatorial Indian Ocean and enhanced trade winds across the equatorial Pacific
- Suppressed convection in the western Indian Ocean and eastern Pacific

The crucial difference—and what justifies the "La Ola" terminology—is that these MJO anomalies propagate eastward around the equator in about 30-60 days, while “La Niña” represents a quasi-stationary coupled ocean-atmosphere state that persists for months to years. In essence, “La Ola-MJO” can be understood as an eastward-traveling, intraseasonal version of the “La Niña”-like circulation pattern, with "La Ola" capturing this essential characteristic of circum-propagation around the earth.

This physical similarity provides a powerful conceptual and metaphorical anchor for public understanding: "If you know “La Niña” brings more rain to Indonesia and drier conditions to Peru, “La Ola-MJO” creates similar patterns—but they move eastward like a wave circling the equator (like a human wave propagating around a stadium, typically called “Ola”) every month or two, rather than staying in place for a year or more."

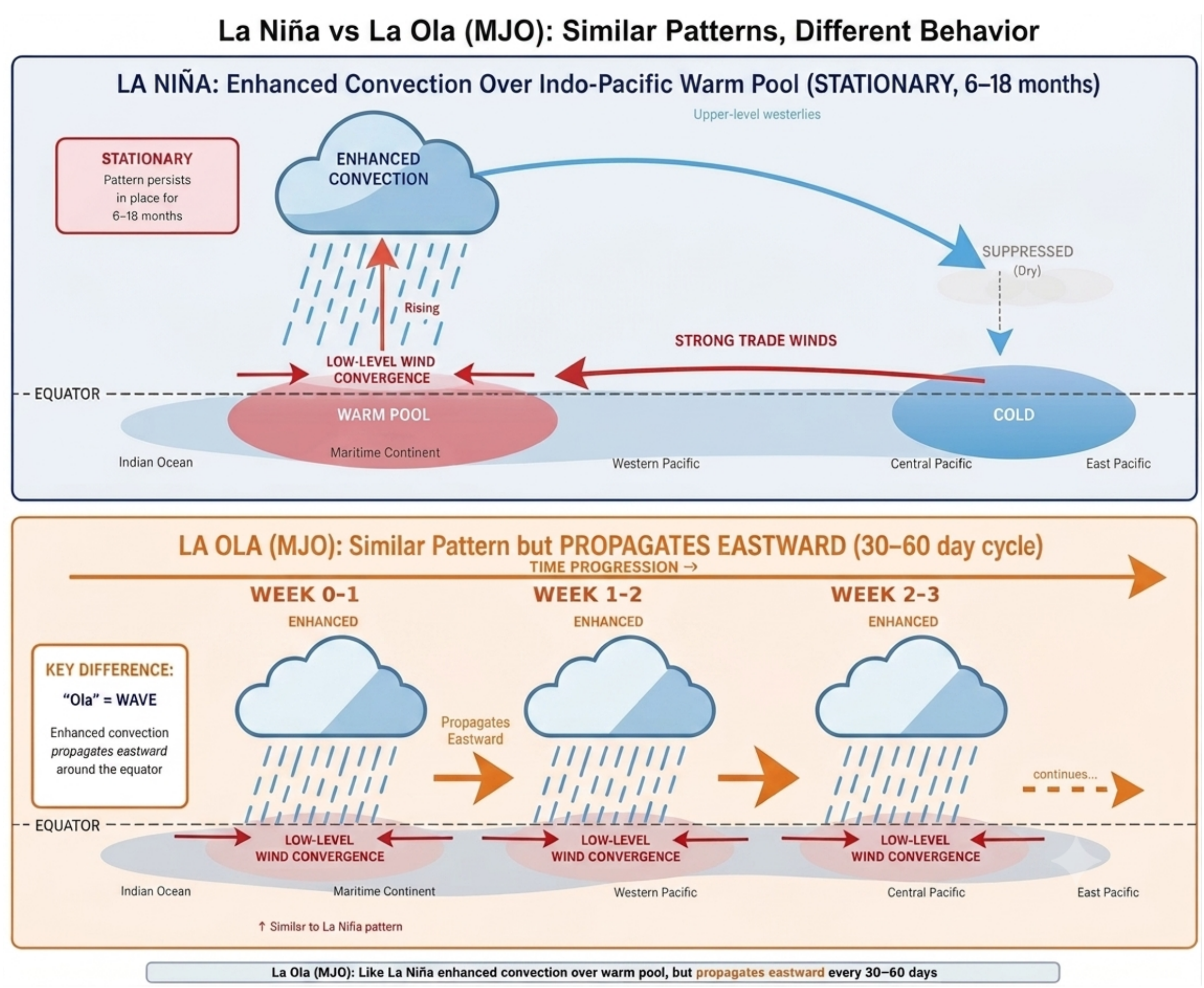

**Fig. 1**. Schematic illustration of “La Nina” (upper) and “La Ola-MJO” (lower). The latter being atmospherically like a “La Niña” (cf. Fig. 2) going around the earth, the phonetical similarity with “La Ola” is physically relevant.

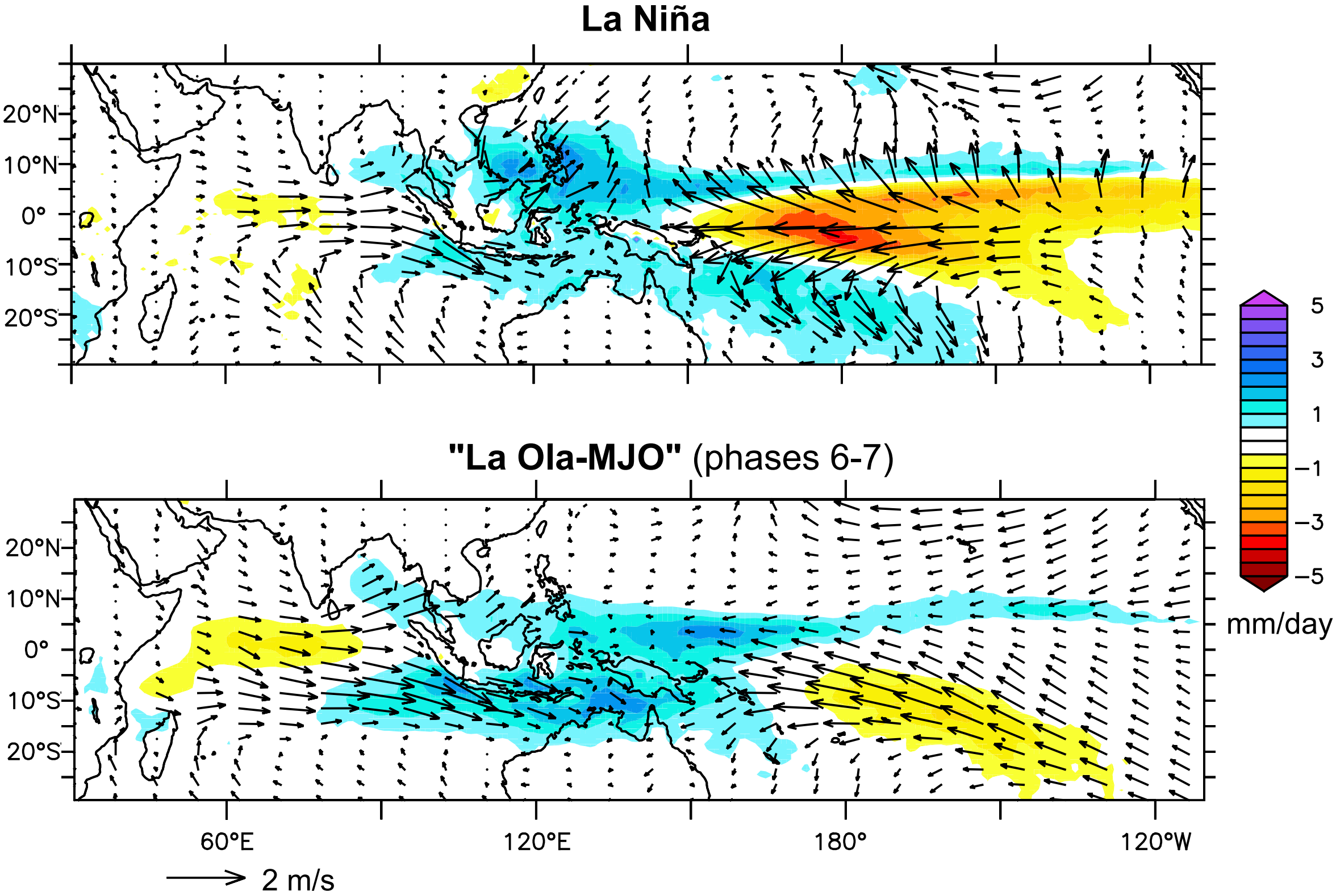

**Fig. 2.** Typical precipitation (color shading, mm day$^{-1}$) and surface wind (vectors, m s$^{-1}$) anomalies associated to “La Niña” (upper panel), and to “La Ola-MJO” (lower panel) when active over the maritime continent (i.e. MJO phases 6-7), in their peak season, boreal winter (November to April), from a regression respectively on -Nino3.4 RSST[9] (Relative Sea Surface Temperature) and RMM1[10] (Real-time Multivariate MJO index 1) normalized indices (cf. Supplementary Material). The “La Ola-MJO”, in its main active phase over the Maritime continent, has qualitatively similar precipitation and wind patterns to “La Niña”, with increased precipitation over the maritime continent, equatorial easterly anomalies in the Pacific and westerlies in the Indian Ocean.

# Why “La Ola-MJO”?

We suggest “La Ola-MJO” offers several advantages as a public-friendly term while maintaining acknowledgment of Madden and Julian's foundational work:

**Honors the original discovery:** By retaining “MJO”, the terminology continues to recognize Roland Madden and Paul Julian's pioneering 1971 discovery (we had brainstormed on other alternatives such as “La MJ-Ola”, but preferred “La Ola-MJO”). This parallels how “El Niño-Southern Oscillation (ENSO)” maintains the connection to both the regional phenomenon and Walker's Southern Oscillation.

**Follows the successful “El Niño/La Niña” convention:** Using Spanish maintains linguistic consistency with the established “El Niño/La Niña” ENSO terminology, immediately signaling to audiences that this is a related but distinct tropical climate phenomenon. The gender difference (“La” vs “El/La”) creates natural distinction while maintaining the familial connection and the atmospheric similarity to the “La Niña” spatial pattern. The “La Ola” term also has good cross-linguistic accessibility. While rooted in Spanish (the world’s second language with most native speakers), the term “La Ola” also means the stadium’s human wave in several other languages such as French, German, Italian, Portuguese, facilitating adoption across multiple language communities. The word is also phonetically simple and easy to pronounce, notably in English.

**Scientifically accurate:** The “La Ola” nickname captures the MJO's most distinctive characteristic, its wave-like eastward propagation around the earth along the equator, as aforementioned. Unlike ENSO, which involves ocean-atmosphere coupling that persists in place for months, the MJO is fundamentally a traveling atmospheric “Ola” of enhanced and suppressed tropical deep convection, related to equatorial wave dynamics and interaction with moisture, convection and boundary layer, typically completing its journey around the globe every 30-60 days. This “La Ola” metaphor could also help explain to the public more specific features of the MJO. For instance, the MJO-related precipitation anomalies reduce strongly away from the Indo-Pacific warmpool: somewhat like a ”La Ola” human wave in areas of the stadium with less public, the MJO completes about half of its circumnavigation around the globe as a quiet, weakened wave that becomes less visible (in terms of rainfall) in equatorial regions with less atmospheric convection, such as Africa, but ‘survives’ and still propagates (with still significant anomalies in e.g. low-level moisture impacting humid heatwaves likelihood[4]).

**Visual and memorable:** “La Ola” evokes an immediate mental image (often seen in the media during World Cups), a wave of atmospheric perturbations moving across the tropical oceans. This visualization is both scientifically sound and intuitively graspable, making it ideal for graphics, animations, and verbal explanations.

**Appropriate scope:** The name is neither too narrow (it does not tie the phenomenon to a specific region) nor too broad (it is clearly distinct from other atmospheric phenomena). It accurately represents the MJO as a large-scale propagating system.

# An invitation for community discussion

To invite the scientific and media communities to discuss about a public-friendly nickname for the MJO, here are some thoughts for brainstorming. Keeping in mind that our goal is not to rename the MJO (we do not want to replace the MJO terminology), but to add a pedagogical metaphor to the MJO that is easy to remember for the public, has some phonetical resemblance to “La Niña”, and that has two physically-sound meanings, as a wave and as a human wave going around a stadium. We recognize that introducing a new term as a nickname alongside an established scientific acronym requires careful consideration. Several factors could help address and mitigate such potential concerns:

**Preserving scientific attribution**: By maintaining the MJO acronym, the terminology continues to honor Madden and Julian's discovery. This approach parallels other successful scientific naming conventions where both historical and descriptive names coexist (e.g., “El Niño” and “Southern Oscillation” merged to form “ENSO”). The scientific community could continue using MJO while “La Ola-MJO” gains traction in public communication, similar to how technical and public terminology naturally diverge in other fields.

**Regional variation**: While Spanish-speaking regions – and English-speaking regions used to Spanish words such as “El Niño/La Niña” – may readily adopt “La Ola-MJO”, regions with other languages might prefer other formulations or maintain MJO. Such regional variation could be natural and acceptable—what matters is that each community has an accessible way to discuss the phenomenon with their stakeholders.

**Distinction from other waves**: While Spanish meteorology uses “ondas tropicales” for Atlantic tropical waves and “ondas ecuatoriales” for equatorial wave types (Kelvin, Rossby), “La Ola” (with the definite article and capital letters) might be established as specifically referring to the MJO, just as “El Niño” specifically means the warm phase of ENSO rather than any warm ocean conditions.

The climate community has successfully demonstrated that accessible terminology accelerates the translation of scientific knowledge into actionable information. El Niño's widespread recognition has enabled sophisticated public discussions about climate impacts and improved preparedness for extreme events. The MJO deserves similar recognition, and merits broader discussion.

We invite feedback and suggestions from:

- **Researchers** on whether “La Ola-MJO” appropriately balances accessibility with scientific attribution, and how it might be introduced in abstracts, textbooks and press releases
- **Operational centers** on the potential benefits and challenges of testing “La Ola-MJO” in public forecasts, and gathering user feedback on comprehension
- **Media outlets** on whether this terminology could facilitate reporting on subseasonal climate variability
- **Educators** on how to effectively teach both the scientific foundation (MJO) and public-friendly terminology (“La Ola”) in complementary ways
- **Regional climate centers** and **WMO** (World Meteorological Organisation) on how this proposal aligns with international coordination efforts in subseasonal forecasting

The science of subseasonal prediction has advanced considerably, with MJO forecasts now extending to 3-4 weeks with meaningful skill, with even further progress expected thanks to AI (Artificial Intelligence) development[11,12]. We suggest that adding to this phenomenon a more accessible nickname, while preserving its scientific attribution, could help reduce the gap between scientific capability and public benefit. “La Ola-MJO” represents an opportunity to discuss how we can better democratize access to life-saving forecast information across ‘our’ warming world.

We welcome comments and suggestions from the community on this proposal.

**Acknowledgements and authors contributions**: All scientific ideas, including the original idea of using “La Ola” as a public-friendly nickname for the MJO (which came out on February 11th, 2026 in our laboratory at the University of French Polynesia from a brainstorming between T.I., B.P. and M. Z.), hypotheses, analyses, and interpretations presented in this work were conceived and developed exclusively by the human authors in our laboratory at UPF. T.I., B.P. and C. R. conceived the schematics in Fig. 1 with the help of a large language model. T.I. did the statistical analysis of Fig. 2. All authors contributed to the concept and ideas, that were first presented to a large audience at the CLIVAR workshop in Tokyo and at the JpGU-AGU joint meeting in Chiba in May 2026. The authors acknowledge the use of NOAA PyFerret software and Python for Figure 2.

**Funding Declaration**: TI is funded by IRD. BP is funded by the SPC (South Pacific Community) through the 'Climate Flagship' project. CR, MZ and MH are funded by the UPF. VS is funded by the PPR Océan et Climat (AI Océan). MC is funded by the ZMT.

**Data Availability Declaration**: All datasets used in this study are open-source and available online. The ERA5 data were downloaded from the Copernicus Climate Data Store and NOAA OISST v2.1 from https://www.ncei.noaa.gov/products/optimum-interpolation-sst. NOAA PyFerret software and Python are open-source, and scripts are available on demand to TI.

**Competing interests**: The authors declare no competing interests.

**Correspondence**: Email should be addressed to takeshi.izumo@ird.fr

**Supplementary Material for:**

# "La Ola-MJO": a public-friendly nickname for the Madden-Julian Oscillation

Takeshi Izumo, Bastien Pagli, Claire Rocuet, Mayalen Zubia, Marania Hopuare, Vateanui Sansine, Maxime Colin

## Data and Methods

Here we used standard observations and reanalysis datasets, indices and statistical methods to compute Figure 2.

The precipitation and surface wind (at 10m) datasets are taken from the ECMWF ERA5 atmospheric reanalysis (Hersbach et al. 2020, 1979-2022), re-gridded through averaging on a 1° grid. To compute the ENSO index, we use Optimum Interpolation SST OISSTv2.1 based on *in situ* observations and satellite measurements (September 1981-December 2024; Reynolds et al. 2002).

The ENSO index used is Niño3.4 (5°N-5°S, 170°W-120°W) relative Sea Surface Temperature (RSST, i.e. SST minus its 20°N–20°S tropical mean), as recommended by Izumo et al. (2020), Van Oldenborgh et al. (2021) and L'Heureux et al. (2024; however we do not use a seasonally-dependent variance adjustment method as in the two latter studies; here we use the simplest – and most justified physically – index: Niño3.4 RSST). RSST-based indices for ENSO are preferable to SST-based ones (and have been adopted recently by the main operational centers), notably in the case of global warming (climate change context) or global cooling (after large volcanic eruptions, cf. Khodri et al. 2016), because atmospheric tropical deep convection interannual anomalies are rather related to RSST than to SST.

The MJO index used is the RMM1 (Real-time Multivariate MJO index; Wheeler and Hendon 2004, downloaded from the Australian Bureau Of Meteorology website), bandpass filtered with a 20-200d Lanczos filter to focus on MJO intraseasonal timescale.

Here we use standard statistical methods. The seasonal cycle (computed by averaging each day/month of the year over the whole period) is removed. For the ENSO analysis based on monthly data, intraseasonal noise is filtered out by a 3-point Hanning filter, so that periods lower than ~2-3 months are removed. For the MJO analysis, high-frequency noise is filtered out by a 3-point Hanning filter on 5-day averaged data, so that periods lower than 10-15 days are removed. For Figure 2, we do a regression of ERA5 precipitation and wind either on -Niño3.4 RSST index for the ENSO analysis (with a minus sign by convention to show the typical anomalies related to a La Niña to ease the comparison with MJO, vice versa for an El Niño), or on RMM1 (Real-time Multivariate MJO index 1) index for the MJO analysis. We use RMM1 rather than RMM2 (not shown), to show the typical anomalies related to MJO active phase over the maritime continent, i.e. between phases 6 and 7 in the RMM phase diagram. Both Niño3.4 RSST and RMM1 indices are normalized by their standard deviation (STD, computed over the season and period considered here), so that the regression coefficients shown in Figure 2 represent the typical anomalies of precipitation and wind, for a typical ENSO or MJO event of 1 STD. The amplitudes can thus be compared quantitatively in Figure 2. The regression is focused on the boreal winter extended season (November to April), which is the peak season of both phenomena. Note that here we do not distinguish the MJO from the boreal summer intraseasonal oscillation (BSISO), as they have fundamentally the same underlying dynamics. Statistics are robust thanks to a sufficiently-large number of effective degrees of freedom, the regressions being done over more than 40 years.

**References for Supplementary Material**